\documentclass[]{aa}
\usepackage{graphicx}
\usepackage{aalongtable}
\begin{document}
\title{Evidence from stellar rotation of enhanced disc dispersal: (I)\\
The case of the triple visual system BD$-$21\,1074 in the  $\beta$ Pictoris association}
\author{S.\,Messina\inst{1}, B.\,Monard\inst{2}, K.\,Biazzo\inst{1}, C.H.F.\,Melo\inst{3}, A.\,Frasca\inst{1}
}
\offprints{Sergio Messina}
\institute{INAF-Catania Astrophysical Observatory, via S.Sofia, 78 I-95127 Catania, Italy \\
\email{sergio.messina@oact.inaf.it}
\and   
Klein Karoo Observatory,   Calizdorp, Western Cape, South Africa
\and   
ESO - European Southern Observatory, Alonso de Cordova 3107, Vitacura Casilla 19001, Santiago 19, Chile \\
}

\date{}
\titlerunning{Rotation of the $\beta$ Pic member BD$-$211074}
\authorrunning{S.\,Messina et al.}
\abstract {The early stage of stellar evolution is characterized by a magnetic coupling between a star and its accretion disc, known as a star-disc locking mechanism. The disc-locking prevents the star to spin its rotation up, and its timescale depends on the disc lifetime, which should not be longer than about 10 Myr. Some mechanisms can significantly shorten this lifetime, allowing a few stars to start spinning up much earlier than other stars and increasing the observed rotation period dispersion among coeval stars.}
{In the present study, we aim to investigate how the properties of the circumstellar environment  can shorten the disc lifetime,  more specifically the presence of a close stellar companion.} {We have identified a few multiple stellar systems, composed of stars with similar masses, which belong to associations with a known age. Since all parameters that are responsible for the rotational evolution, with the exception of environment properties  and initial stellar rotation,  are similar for all components, we expect that significant differences among the rotation periods can only arise from differences in the disc lifetimes.   A photometric timeseries allowed us to measure the rotation periods of each component, while high-resolution spectra provided us with the fundamental parameters, $v\sin{i}$ and chromospheric line fluxes.} {In the present study, we have collected timeseries photometry of BD$-$21\,1074,  a member of the 21-Myr old $\beta$ Pictoris association,   and measured  the rotation periods of its brightest components A and B. They differ significantly, and the component B, which has a closer companion C,  rotates faster than the more distant and isolated component A.   It also displays a slightly higher chromospheric activity level.  } {Since components A and B have similar mass, age, and initial chemical composition, we \ can ascribe the rotation period difference to either different initial rotation periods or different disc-locking phases arising from the presence of the close companion C. 
In the specific case of BD$-$21\,1074, the second scenario seems to be more favored. However, a statistically meaningful sample is yet needed to be able to infer which scenario is more likely. In our hypothesis of different disc-locking phase,  any planet orbiting this star, if found by future investigations, is likely formed very rapidly owing to a gravitational instability mechanism, rather than core accretion.    Only  a large  difference of initial rotation periods alone could account for the observed period difference, leaving comparable disc lifetimes.  }
\maketitle
\rm

\section{Introduction}
The early stage evolution of low-mass stars is characterized by a magnetic coupling between the accretion disc and the star's convective envelope (see, e.g., M\'enard \& Bertout \cite{Menard99}). The central star gains angular momentum from the infalling gas, but this excess angular momentum is, on the other hand, dissipated by accretion-driven winds (Matt \& Pudritz \cite{Matt05}, \cite{Matt08a}, \cite{Matt08b}) and mass ejection episodes caused by magnetospheric reconnection (Zanni \& Ferreira \cite{Zanni13}). As a result, the star's rotation period remains about constant (Bouvier et al. \cite{Bouvier93}; Edwards et al. \cite{Edwards93}; Rebull et al. \cite{Rebull04}) as far as the star-disc interaction is effective, despite the ongoing stellar radius contraction. This coupling mechanism is commonly referred to as 'star-disc locking' (see, e.g., Camenzind \cite{Camenzind90}; Koenigl \cite{Koenigl91}; Shu et al. \cite{Shu94}). Once the disc, or at least the inner and denser part of it, has been dissipated, the star-disc locking is no longer effective, and the external star's envelope can freely spin its rotation up, owing to stellar radius contraction and angular momentum conservation.\\
\indent
The disc lifetime and the timescale of star-disc locking are not easily measured. Even if the dust can be detected by infrared observations, the cold gas, which represents the dominant fraction of disc mass, is hard to detect. Data retrieved from infrared surveys, which map the dust component,  show that the frequency of low-mass stars with discs decays exponentially with a decay time of $\tau$=2--3 Myr for the inner discs, and $\tau$=4--6 Myr for primordial discs (Ribas et al. 2014),  or even longer if we take for granted the revised ages of star-forming-regions by Bell et al. (\cite{Bell13}).   No primordial discs are detected at ages older than 10--20 Myr.
The presence of accreting gas and, thus, of star-disc locking still in operation, can be indirectly detected by signatures of accretion, such as strong and broad emission lines and a very large photometric variability arising from the infalling gas (M\'enard \& Bertout \cite{Menard99}).
Observational studies  carried out by Jayawardhana et al. (\cite{Jayawardhana99}, \cite{Jayawardhana06}) show that the disc frequency as probed by  accretion signatures also decreases exponentially with no detection at ages older than about 8 Myr.  A very low fraction of accretors (2--4\%) has been found in the 10-Myr old \object{$\gamma$ Velorum} association (Frasca et al. \cite{Frasca14}).  
These studies allow us to put an upper limit to the disc's lifetime and to the timescale of effective star-disc locking
at about 8--10 Myr. An important point that we can infer from the frequency distributions is that a fraction of stars has no disc already at an age of about 1 Myr. Therefore, the disc dissipation likely has a variety of mechanisms, and a few of them are more effective than others. This circumstance also reflects in the rotation period distribution.
Focusing on solar-mass stars, we observe that the period distributions, although peaked, have a dispersion already at an age of 1--3 Myr (\object{ONC}, Herbst et al. \cite{Herbst01}). They become increasingly larger at 2--4 Myr (\object{NGC2264}, Lamm et al. \cite{Lamm05}), which is still larger in the  6-Myr \object{$\epsilon + \eta$ Chamaeleontis} and in the  8-Myr  \object{TW Hya} associations (Messina et al. \cite{Messina10}, \cite{Messina11}). This evidence confirms that the disc's lifetime has a range of values, and a few stars start spinning up earlier/later than other stars.\\
\indent
The disc lifetime can depend on different parameters. For example, Collier Cameron et al. (\cite{CollierCameron93}, \cite{CollierCameron95}) find a dependence  on the disc mass. In the case of massive discs, the ZAMS rotation is almost independent on both mass disc and initial angular momentum. In the case of low-mass discs, the star-disc locking is never very effective and the star arrives onto the ZAMS with a greater fraction of its initial angular momentum. The disc lifetime can also depend on the properties of circumstellar environment. The formation of a close-in planet contributes to clearing out the inner disc, while a close stellar companion can strongly speed up the whole disc dispersal.  The disc lifetime and disc-locking timescales are considerably shortened in both cases. \\
\indent
Our aim is to contribute to a better understanding of the dependence of the disc lifetime on the circumstellar environment, and specifically, on the presence of close companions. The best targets to study such dependence are systems where all other parameters that may affect the disc lifetime are fixed. Such targets are represented by multiple stellar systems belonging to associations/clusters of known age and consist of equal mass components. In this case, initial angular momentum, age, metallicity, and disc mass are expected to be similar for all components, and any variation in disc lifetime can be ascribed to differences in the stellar environments.\\
\indent
A different disc lifetime and star-disc locking timescale affects the stellar rotation periods, since the component with the shorter disc-locking time is expected to start spinning up earlier than the other. Indeed, we search for differences in the rotation periods among the system's components as a probe for enhanced disc dispersal.\\
\indent
As a part of the RACE-OC project (Rotation and ACtivity Evolution in Open Clusters; Messina \cite{Messina07}; Messina et al. \cite{Messina08}, \cite{Messina10}) aimed to search for the rotation periods of members of young stellar associations, we have identified a number of such multiple systems. In this paper, we present the results for the multiple system \object{BD$-$21\,1074}, which is member of the young \object{$\beta$ Pictoris} association.
In Sect. 2, we present the available information from the literature on the selected target. We describe our photometric and spectroscopic observation in Sect. 3, their analysis in Sect.4, a discussion of the results in Sect.5, and our conclusions in Sect. 6. 
 \cite{}

\section{The BD$-$211074 system}
\begin{figure}
\begin{minipage}{9cm}
\centerline{
\includegraphics[width=90mm,height=90mm,angle=0]{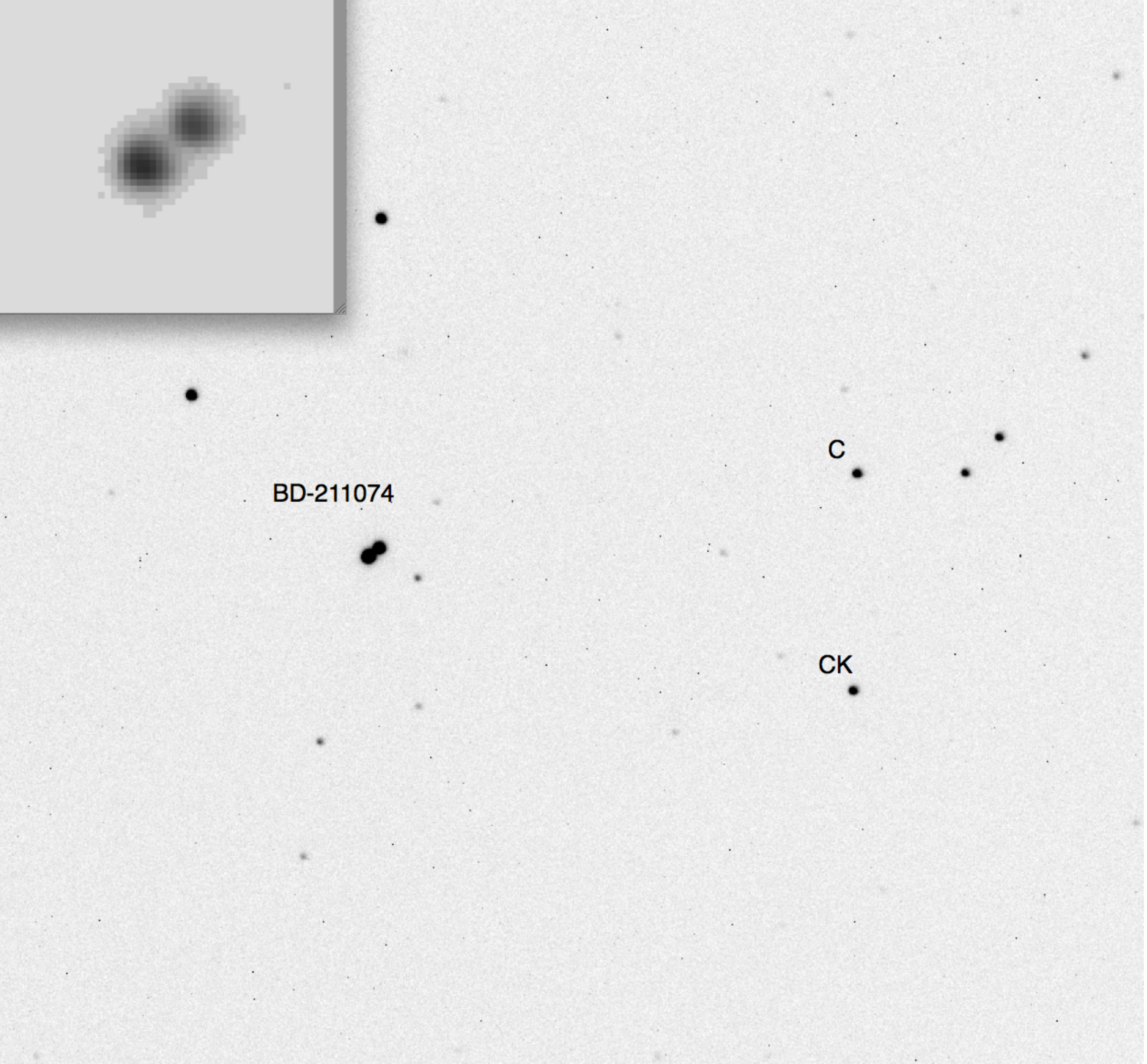} 
}
\end{minipage}
\caption{\label{fov}Portion of the field of view of the  \object{BD$-$21\,1074} system imaged in the R-band in a night with the average 2$\arcsec$ seeing (North is up and east is to the left). The top-left panel displays a 
magnified view of the system, where the A component is at bottom left, whereas the unresolved components BC are at top-right.}
\vspace{0cm}
\end{figure}

The object \object{BD$-$21\,1074} is reported in the Washington Catalog of Visual Double Stars (Mason et al.  \cite{Mason01}) as a stellar system consisting of three visual M-type stars. The primary component, \object{BD$-$21\,1074A} (= \object{GJ 3331} = \object{DONNER 93A} = \object{TYC 5913\,452\,1} =  \object{2MASS J05064991$-$2135091}), is an M1.5 star with V = 10.29 mag. It is at an angular distance of 8.2${\arcsec}$ from the secondary component, \object{BD$-$21\,1074B} (\object{GJ 3332} =  \object{DONNER 93B} =  \object{TYC 5913\,1376\,1} = \object{2MASS J05064946$-$2135038}), an  M2.5 star with V = 11.67 mag (see Fig.\,\ref{fov}).
The tertiary component, \object{BD$-$21\,1074C} (= \object{DONNER 93C}) has  an angular separation of 0.8${\arcsec}$  from the secondary component and magnitude V = 12.67 mag with a M5 or later spectral type.\\
\indent
For this system, wide band photometry of all components is available, which shows evidence for long-term variability.
Weis (\cite{Weis91}) reports the following: V = 10.29 mag and the colors B$-$V = 1.51,  V$-$R = 1.09, and R$-$I = 0.90 mag for the A component; and a combined magnitude and colors of V = 11.30 mag  and  B$-$V = 1.66, V$-$R = 1.32,  R$-$I = 1.14 mag for the  BC component. Reid et al.  (\cite{Reid04}) report V = 10.28 mag for the primary, V = 11.67 mag for the secondary, and V = 12.67 mag for the tertiary component. Subsequent differential observations of the BC minus A components obtained by Jao et al. (\cite{Jao03}) in the years 1999--2002  exhibit significant differences with respect to the Weis (\cite{Weis91}) measurements: $\Delta$V changed from 1.01 to 0.722 mag, $\Delta$ (V$-$R) from 0.23 to 0.15, and $\Delta$ (R$-$I) from  0.24 to 0.35 mag. Riedel et al. (\cite{Riedel14}) report the following deblended magnitude V = 10.41 (V$-$I = 2.16), V = 11.44 (V$-$I = 2.58), and V = 12.44 (V$-$I = 2.71) mag for the A, B, and C components, respectively.\\
\indent
Evidence of orbital motion of the A with respect to the B component are marginal: the separation ($\rho$) changed from 8.6${\arcsec}$ in 1920 to 8.22${\arcsec}$ in 1991.7 to  8.2${\arcsec}$ in 2001 to 8.3${\arcsec}$ in 2008.96, whereas the position angle (PA) changed from 323$^\circ$ to 312$^\circ$ to 229.3$^\circ$ to 311$^\circ$, respectively  (see Mason et al.  \cite{Mason01}; Fabricius et al.  \cite{Fabricius02}; Riedel et al.  \cite{Riedel14}). The changes of the B with respect to the C component  are much more significant: $\rho$ passed from 1.2${\arcsec}$ in 1930 to 0.76${\arcsec}$ in 2008 to 0.8${\arcsec}$ in 2010, whereas PA from 30$^\circ$ to 329$^\circ$ to 148.2$^\circ$ to 321$^\circ$, respectively (see Mason et al.  \cite{Mason01}; Tokovinin et al.  \cite{Tokovinin10}; Riedel et al.  \cite{Riedel14}).\\
\indent


\begin{table}
\caption{\label{obslog}Log of the photometric observations.}
\begin{tabular}{l l l l  l}
\hline
Date & Tel. & obs. & Filter &  Bin.\\
d/m/y & aper (cm) & \# & &\\
 \hline
06/12/12   &30 &        772 & R &  2$\times$2\\
11/12/12   &30 &        442 & R &   2$\times$2\\
16/12/12   &35 &        282 & R & 2$\times$2\\
           &35 &        282 & V & 2$\times$2\\
	   &35 &        282 & I & 2$\times$2\\
17/12/12   &30 &	182 & R & 2$\times$2\\
	   &35 &	 95 & V & 2$\times$2\\
18/12/12   &35 &       372  & R &2$\times$2\\
	   &35 &	7   & V &2$\times$2\\
	   &35 &	7   & I &2$\times$2\\
           &30 &       656  & R & 2$\times$2\\
26/12/12   &30 &       714  & R &  1$\times$1\\
28/12/12   &30 &       405  & R &  1$\times$1\\
29/12/12   &30 &       498  & R &  1$\times$1\\
02/01/13   &30 &       395  & R &  1$\times$1\\
06/01/13   &30 &        66  & R &  1$\times$1\\
19/01/13   &30 &       101  & R &  1$\times$1\\
21/01/13   &30 &       219  & R &  1$\times$1\\
22/01/13   &30 &       264  & R &  1$\times$1\\
23/01/13   &30 &       273  & R &  1$\times$1\\
24/01/13   &30 &       312  & R &  1$\times$1\\
26/01/13   &30 &       213  & R &  1$\times$1\\
27/01/13   &30 &       199  & R &  1$\times$1\\
28/01/13   &30 &        88  & R &  1$\times$1\\
\hline
\end{tabular}
\end{table}

The \object{BD$-$21\,1074} system is an EUV, EUVE, and X-ray source. It was detected in the extreme ultraviolet (EUV) by the ROSAT Wide Field Camera (WFC) and listed in the WFC Bright Source Catalogue as \object{RE 0506-213} (Pounds et al.  \cite{Pounds93}). Subsequently, it was also detected by the Extreme Ultraviolet Explorer (EUVE) in the all sky survey (\object{EUVE 0506-215}; Bowyer et al.  \cite{Bowyer94}) and in the EUVE pointed observations (Christian et al.  \cite{Christian98}), which, however, did not detect any evidence for variability. Finally, it was also detected at X-ray wavelengths by ROSAT and reported as \object{1RXS\, J050649.5-213505} by Voges et al. (\cite{Voges99}) with Log(L$_{\rm X}$/L$_{\rm bol}$) = $-$3.24 (Riedel et al.  \cite{Riedel14}). \\
\indent
Both components, A and the unresolved BC, are listed in the catalog  of UV Cet-type flare stars compiled by Gershberg et al. (\cite{Gershberg99}). 
For this system, measurements of Li abundance and  equivalent widths of some activity indicators are available. da Silva et al. (\cite{daSilva09}) report EW$_{\rm Li}$ = 20 m\AA , A$_{\rm Li}$ = $-$0.88, and T$_{\rm eff}$ = 3613 K for \object{BD$-$21\,1074A} and EW$_{\rm Li}$ =  20 m\AA , A$_{\rm Li}$ =  $-$1.07, and T$_{\rm eff}$ = 3496 K for BD$-$21\,1074BC. Gizis et al. (\cite{Gizis02}) report  EW$_{\rm H\alpha}$ = 2.413 \AA \,and 5.775 \AA \,for the primary and the unresolved secondary/tertiary components, respectively, whereas Riedel et al. (\cite{Riedel14}) find  EW$_{\rm H\alpha}$ = 2 \AA \,and 5.2 \AA \,for the primary and the unresolved secondary/tertiary components, 1.10 and 1.13 for the Na index (see Lyo et al.  \cite{Lyo04}), and EW$_{\rm K_I}$ = 0.7\,\AA\,\, and EW$_{\rm K_I}$ = 0.8\,\AA, respectively.  Strong $\ion{Ca}{ii}$ H\&K emission is reported by Bidelman (\cite{Bidelman88}).  Reiners et al. (\cite{Reiners12}) report L$_{\rm H\alpha}$/L$_{\rm bol}$ = $-$3.89 and provide the only available measurement of the projected rotational velocity of the primary component ($v \sin{i}=$  5.3 kms$^{-1}$).

 \begin{figure*}
\begin{minipage}{18cm}
\centerline{
\includegraphics[width=140mm,height=190mm, angle=90]{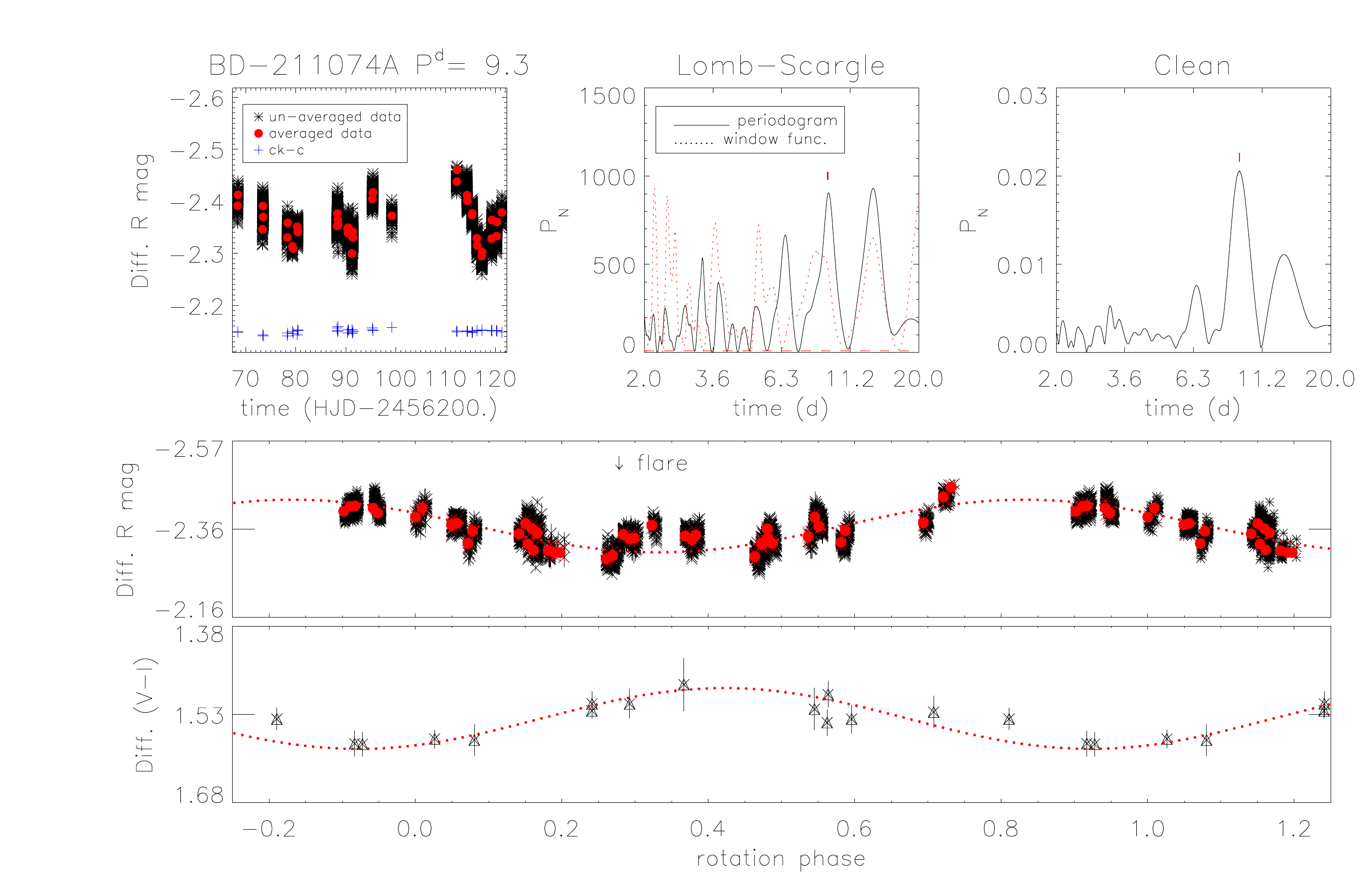}
}
\end{minipage}
\caption{\label{bd-211074A}Results of the rotation period search for \object{BD$-$21\,1074A}. \it Top panels\rm:  differential magnitude time series of BD$-$21\,1074A and of CK$-$C (arbitrarily shifted in magnitude), where
asterisks and bullets represent the un-averaged and averaged (2hr bin) data, respectively;  L-S and Clean periodograms showing the power peak corresponding to the rotation period.  \it Middle panel\rm:  phased light curve with the rotation period with the sinusoidal fit.  \it Bottom panel\rm: differential (V$-$I) color curve and the sinusoidal fit.}
\vspace{0cm}
\end{figure*}  

\begin{figure*}
\begin{minipage}{18cm}
\centerline{
\includegraphics[width=140mm,height=190mm, angle=90]{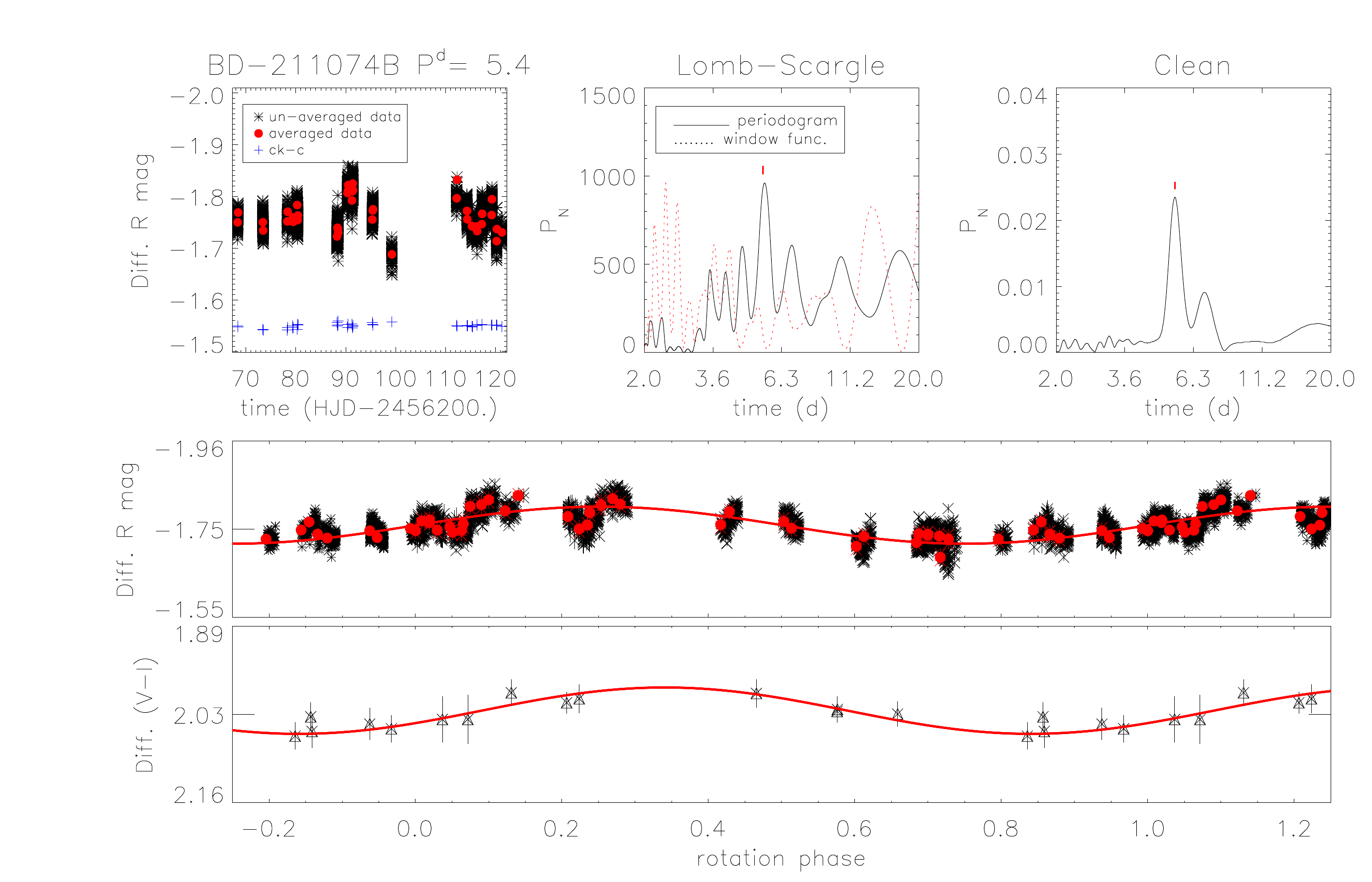}
}
\end{minipage}
\caption{\label{bd-211074B}Results of the rotation period search for BD$-$21\,1074B (see also caption of Fig.\,\ref{bd-211074A}). }
\vspace{0cm}
\end{figure*}  

\section{Observations}
\subsection{Photometry from the literature}
The system \object{BD$-$21\,1074} has been photometrically monitored by ASAS (All Sky Automated Survey; Pojmanski  \cite{Pojmanski97}) since November 2000 and by SuperWASP (Butters et al.  \cite{Butters10}) since September 2006. Owing to the low spatial resolution, these observations could not resolve the components, allowing only  integrated magnitudes. These photometric time series were analyzed by Messina et al. (\cite{Messina10},  \cite{Messina11}) for the period search, and the  rotation periods P = 13.3 d and P=13.4 d were inferred from ASAS and SuperWASP, respectively.  Assuming that the A component, which is brighter by about 1 mag than the BC components, dominates the flux variability, this rotation period was attributed to \object{BD$-$21\,1074A}. 
When this period is combined with the stellar radius (see Sect.4) and the projected rotational velocity,  we find that $\sin{i}$ $\simeq$ 1.16, \rm which is  larger than the maximum value admitted for an inclination $i$ = 90$^{\circ}$. Assuming that both $v\sin{i}$ and stellar radius are correct, the expected maximum rotation period should be   smaller than about 11.5 days.   

To accurately determine the rotation periods of the system's components and to measure their physical parameters and the level of magnetic activity, we planned photometric and spectroscopic observations. Photometry was collected with a small-aperture telescope but sufficiently large  to spatially resolve the component A from BC, and spectroscopy was obtained with the HARPS spectrograph at ESO.
\subsection{New photometry}

The object \object{BD$-$21\,1074} was observed during 18 nights (see Table\,\ref{obslog}) from December 6, 2012 to January 28, 2013 at the Klein Karoo Observatory (225 m a.s.l, Western Cape, South Africa).  It was observed with a 30cm (f/8) RCX-400 telescope with a 21$\arcmin$$\times$14$\arcmin$ field of view, a scale of 0.82$\arcsec$/pixel,  and the SBIG ST8-XME CCD camera and BV(RI)$_c$ filter set. In two nights, observations were also collected with a 35cm (f/8) RCX-400 telescope equipped with the same camera and filters. We could collect a total of 6453 frames in the R filter, 108 in the V filter during three nights, and 13 frames in the I filter during two nights. In December 2012, the observations were collected with a 2$\times$2 CCD binning, whereas  we adopted an unbinned mode to optimize the spatial resolution in the subsequent nights. The data reduction was performed using the IRAF\footnote{IRAF is distributed by the National Optical Astronomy Observatory, which 
is operated by the Association of the Universities for Research in Astronomy, inc. (AURA) under 
cooperative agreement with the National Science Foundation.} tasks within DAOPHOT. Although \object{BD$-$21\,1074A} turned out to be almost spatially resolved from BD$-$211074BC  to prevent any flux contamination by the nearby components, especially in case of poor seeing, we used the Point Spread Function (PSF) photometry technique to extract the magnitudes of all stars detected in our frames. The component BC were unresolved in our observations. The integration time was set to 24 sec in the R filter, and 13 sec  in both the V and I filters. For each night, the observations were performed for about four consecutive hours. \\
After bias subtraction and flat fielding, we extracted a timeseries of magnitudes for each detected star that were cleaned by applying a 3$\sigma$ threshold to remove outliers. In the magnitude range of our variable targets (Var) and comparison stars (C and CK), the nominal R-band photometric accuracy provided by DAOPHOT, which is based only on the photon statistics with 24-sec integration, turned out to be better than 0.005 mag. However, to measure
the effective photometric accuracy of our observations,
instead of using the values provided by DAOPHOT, we sectioned
our timeseries into bins with  2-hr widths (corresponding
on average to about 50 consecutive measurements)
computing means and standard deviations. We found that
the average standard deviation for the Var$-$C measurements
was $\sigma_{\rm Var-C}$ = 0.015 mag, whereas $\sigma_{\rm CK-C}$ =
0.008 mag. These values represent a more effective estimate
of the precision of our photometry. Such values are conservative,
because these comparisons are about 1 mag fainter than the target, and the true observational accuracy could be, in principle, even
better in the case that some substantial variability exists within the timescale
that is close to our fixed binning time interval (i.e. 2 hr).\\
 \indent
In the vicinity of \object{BD$-$21\,1074}, we could select two stars, \object{TYC\,5913\,431\,1} (RA = 05$^h$\,06$^m$\,29$^s$.044; Dec =  $-$21$^{\circ}$\,34$^\prime$\,24$^{\prime\prime}$.15; V = 12.25 mag) and \object{TYC\,5913\,478\,1} (RA = 05$^h$\,06$^m$\,29$^s$.358; Dec = $-$21$^{\circ}$\,36$^\prime$\,33$^{\prime\prime}$.21; V = 12.35 mag), which were found to have stable light during the whole run. Therefore, they turned out to be suitable to serve as comparison (C) and check (CK) stars, respectively, for our differential photometry.
During the observation run, the differential magnitude of the check minus the comparison star remained constant with a mean value of $\Delta$R = $-$0.055 mag and a standard deviation $\sigma$ = 0.008 mag.

\subsection{Spectroscopy}
We observed both components in January $9-12$, 2013 with the HARPS spectrograph 
($R\sim$110\,000, $\lambda$=3800--6900\,\AA; Mayor et al.  2003) at the 
ESO 3.6\,m telescope in La Silla (Chile).  The log book of the spectroscopic 
observations is given in   Table~\ref{tab:spectroscopic_observations}. \\
\indent
From these observations, all Balmer hydrogen lines within the HARPS wavelength range 
(i.e. from the H$\alpha$ to the H$\epsilon$) resulted in emission,  with the 
\ion{Ca}{ii} H\&K  lines and the \ion{He}{i} line at   $\lambda$5876\,\AA. Moreover,   a weak Li\,{\sc i} absorption line at $\lambda$6707.8\,\AA~is 
detected in the spectra of both stars with EWs similar to those reported by da Silva et al. (\cite{daSilva09}) values. The radial velocities (RV) of both components are 
found in good agreement with earlier values (+21.2$\pm$0.9 and +23.7$\pm$1.7 km s$^{-1}$, respectively, 
for the A and BC components), as reported by Gizis et al. (\cite{Gizis02}).

\begin{table}[h]  
\caption{Log of the spectroscopic observations.}
\label{tab:spectroscopic_observations}
\begin{center}  
\begin{tabular}{cccccc}
\hline
\hline
\noalign{\smallskip}
$\alpha$ (J2000)   & $\delta$ (J2000)  &  Date      &  UT      & $t_{\rm exp}$ \\
(h:m:s)     &  (\degr: $^\prime$ : \arcsec)  & (d/m/y)    & (h:m:s)  &  (s)    \\ 
\noalign{\smallskip}
\hline
~\\
\multicolumn{5}{c}{BD$-$21\,1074A}\\
\hline
\noalign{\smallskip}
05:06:49.91 & $-$21:35:09.1 & 10/01/2013 & 03:48:12 & 900  \\ 
05:06:49.91 & $-$21:35:09.1 & 13/01/2013 & 03:39:00 & 900  \\ 
\noalign{\smallskip}
\hline
~\\
\multicolumn{5}{c}{BD$-$21\,1074B}\\
\hline
\noalign{\smallskip}
05:06:49.46 & $-$21:35:03.8 & 10/01/2013 & 03:29:56 & 900  \\ 
05:06:49.46 & $-$21:35:03.8 & 11/01/2013 & 03:21:53 & 900  \\ 
05:06:49.46 & $-$21:35:03.8 & 13/01/2013 & 03:20:38 & 900  \\ 
\noalign{\smallskip}
\hline
\end{tabular}
\end{center}
\end{table}

\section{Analysis and results}

\subsection{Photometric rotation periods}
The un-averaged and averaged R-band photometric time series were analyzed for the rotation period search with the Lomb-Scargle (L-S; Scargle  \cite{Scargle82}) and the Clean (Roberts et al.  \cite{Roberts87}) periodograms. An estimate of the False Alarm Probability was done using Monte Carlo simulations according to the approach outlined by Herbst et al. (\cite{Herbst02}). The uncertainty on the rotation period determination was estimated following the prescription given by Lamm et al. (\cite{Lamm04}). A more detailed description of the steps of our analysis is given in Messina et al. (\cite{Messina10}).\\
The results for \object{BD$-$21\,1074A} and BD$-$21\,1074BC are reported in Figs.\,\ref{bd-211074A}--\ref{bd-211074B}. We also plot the series of CK-C data to show their light constancy during the run interval. In the  top middle and right panels, we plot the L-S and Clean periodograms. In the L-S periodogram we overplot the window function. The normalized power corresponding to a 1\% FAP resulted to be 10 and 12 for the components A and BC, respectively. In the middle panel, we plot the light curve phased with the rotation period, whereas the differential V$-$I color curve phased with the same period is shown in the bottom panel. \\
\indent
In the case of \object{BD$-$21\,1074A}, the L-S periodogram exhibits two major peaks with comparable power. The longer is at P = 13.64 d (FAP $<$ 0.1\%), which  agrees with the values found by Messina et al. (\cite{Messina10},  \cite{Messina11}) and obtained by the integrated flux of all components.  Although this value is still consistent with the stellar radius and projected rotational velocity within the uncertainties, it is not consistent with the star-disc locking an subsequent spinning up scenario, as discussed in Sect.\,5.    The other peak is at P = 9.30 d that is the only one dominant in the Clean periodogram, where the effect of the observation spectral window is removed.  We note a clear anticorrelation between the R  and the V$-$R color curves, 
where the star is fainter when the color is bluer, which is the typical trend observed when the variability has a significant contribution by faculae (see, e.g., Messina  \cite{Messina08}).\\
\indent
In the case of BD$-$211074BC, the magnitude difference between components B and C is $\Delta$V $\simeq$ 1.2 mag, which implies that the luminosity of the component C is more than five times smaller than the luminosity of component B. Therefore, we expect that the variability is dominated by the component B, whose peak in the periodogram should be the most powerful. 
Both the L-S and Clean periodograms show  one major power peak at P =  5.40\,d and a lower, but significant, peak at P = 6.90 d. We attribute the shorter period to the component B, while P = 6.9\,d could be due to the component C, or it is a residual alias. In contrast to the component A, we note a positive correlation between the R and the V$-$I color curves, which is the typical trend observed when the variability is dominated by cool spots.

Both component A and BC are known to be flare stars (Gershberg et al.  \cite{Gershberg99}). Beside the light rotational modulation likely owing to dark spots unevenly distributed across the stellar longitudes, this implies that also transient light increases, as in flares and micro-flares, are expected to be detected. Indeed, we note a significant (larger than the achieved photometric precision) variability during the same night on a time scale of about 1--2 hr. Moreover, during our monitoring, we could observe one such flare (see Fig.\,\ref{flare}) in the R band that exhibited a magnitude increase of 0.06 mag and lasted about half an hour. As shown in Fig.\,\ref{bd-211074A}, it occurred at the phase of minimum light, which is when the star exhibits to the observer the most spotted hemisphere. In the SuperWASP timeseries, we could detect other two flares in the V band: the first occurring at HJD = 2454453 with a magnitude increase of $\sim$ 0.6 mag and a duration of about 2 hr and a second at HJD = 2554489 with a magnitude increase of 0.7 mag that was, unfortunately,  only partially observed.

\subsection{Stellar parameters from spectroscopy}
\begin{figure*}
\begin{center}
\begin{minipage}{18cm}
\centerline{
\includegraphics[width=10.cm]{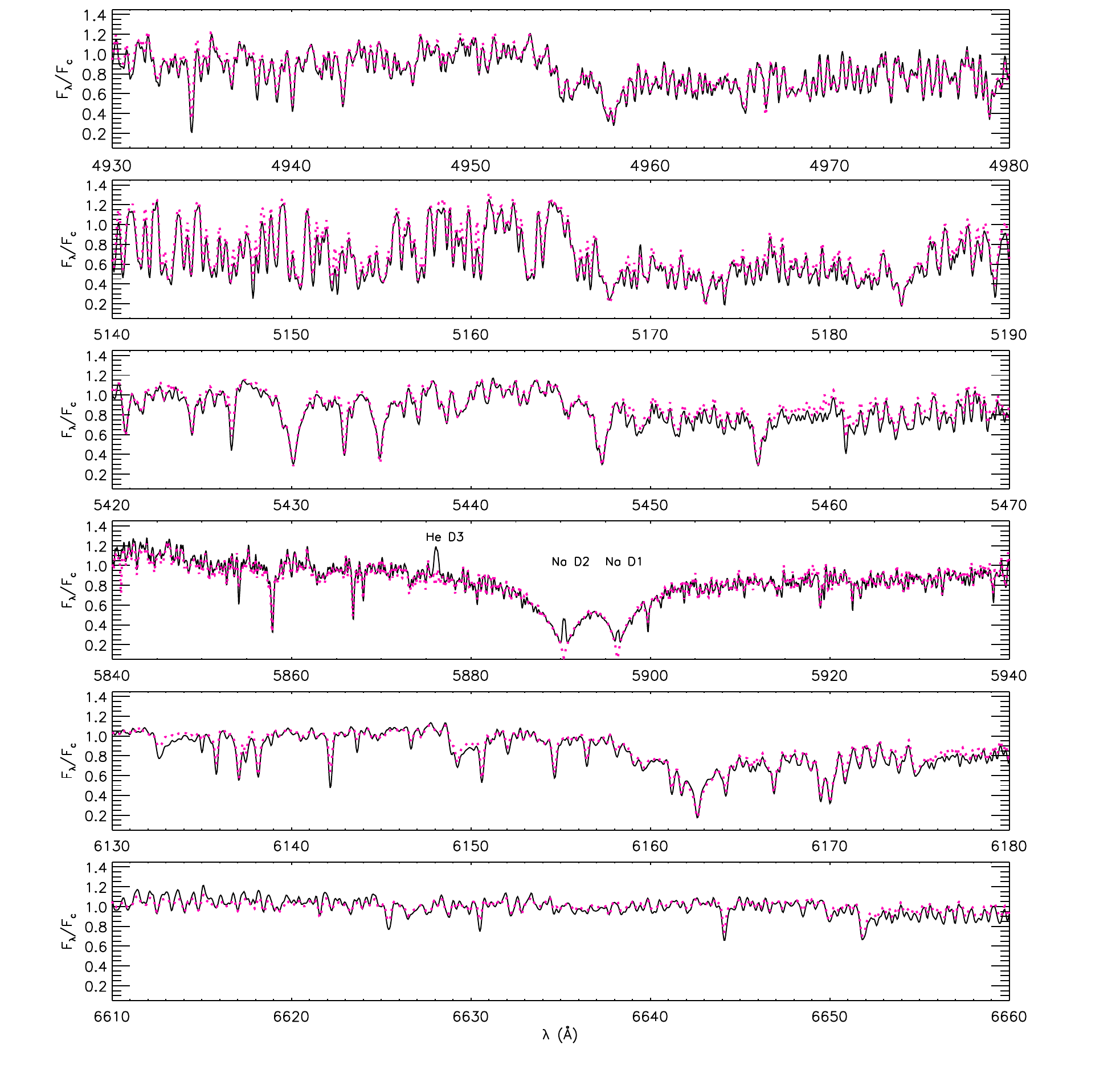}
\includegraphics[width=10.cm]{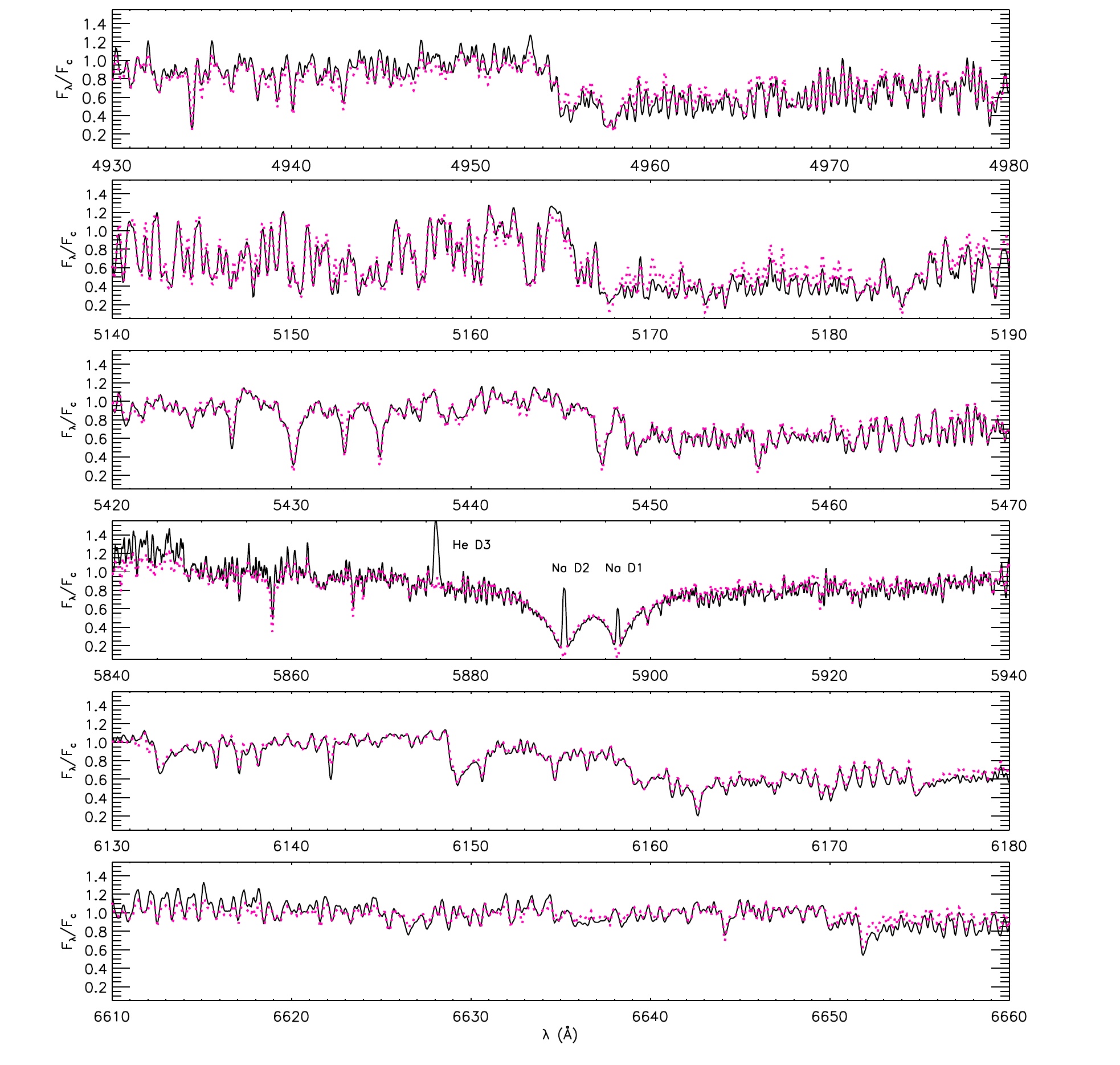}
}
\end{minipage}
 \caption{Portions of the HARPS spectrum of \object{BD$-$21\,1074A} (\it left\rm) and BD$-$21\,1074B (\it right\rm) around some TiO bands and in the 
 \ion{Na}{i} region. The M1.5\,V best template (\object{HD\,36395}) is overplotted with 
 dotted lines.}
 \label{fig:spectrum_bd211074a}
\end{center}
\end{figure*}

Fundamental parameters, such as effective temperature, $T_{\rm eff}$, surface gravity, $\log{g}$, and rotational 
velocity, $v\sin{i}$, for both stars were derived through the ROTFIT code (Frasca et al.  \cite{Frasca03})
 and based on a $\chi^2$ minimization procedure developed in 
IDL\footnote{IDL (Interactive Data Language) is a registered trademark of IT Visual Information 
Solutions.}. This procedure provides us with the best match of the observed spectrum with a grid 
of high-resolution spectra of real   stars. The parameters $T_{\rm eff}$ and $\log{g}$ were derived using spectra 
retrieved from the ELODIE archive\footnote{http://atlas.obs-hp.fr/elodie/} 
(Moultaka et al.  \cite{Moultaka01}) and with well determined parameters (PASTEL catalogue\footnote{http://vizier.u-strasbg.fr/viz-bin/VizieR?-source=B/pastel}; Soubiran et al. 2010). 
The  ELODIE Archive spectra span the wavelength range 4000--6800\,\AA~and have a nominal resolution of $R=42\,000$. 
For this reason, the HARPS spectra of the targets were degraded at the same resolution as ELODIE.
Moreover, as the spectra of the system components are affected by molecular bands, it was difficult to 
perform a homogeneous normalization to the local continuum for both targets and ELODIE templates. 
We thus used a modified version of the code purposely developed for M-type 
stars as in  Alcal\'a et al. (\cite{Alcala11}). In brief,  to exploit the high sensitivity to temperature 
and gravity of molecular bands, we analyzed selected spectral regions, mainly 
around TiO bands ($\lambda\lambda$4950, 5166, 5450, 6158, 6650, 6680\,\AA). 
Then, we normalized each spectral segment with respect to the average stellar flux in a 
window of about 10\,\AA\ blueward of the band-head (see the example reported in Fig.~\ref{fig:spectrum_bd211074a}). 
At the end of this procedure, ten templates that best matched the HARPS spectrum were chosen per each analyzed spectral segment and their fundamental parameters were used to compute 
the weighted averages.
  We used the library of ELODIE templates for deriving $T_{\rm eff}$ and $\log{g}$, because it contains several M-type stars in a suitable range of parameters.
For a best determination of the $v \sin{i}$, we have used the full resolution of the HARPS spectra (R = 110,000)  instead selecting  a few M-type stars observed with the same instrument as templates. For this purpose, the best template for both A and B components of BD$-$21\,1074 is the spectrum of \object{HD\,36395} (M1.5\,V) retrieved from the ESO Archive\footnote{http://archive.eso.org}.
 
The final results are $T_{\rm eff}=3720\pm100$\,K, $\log{g}=4.7\pm0.1$\,dex, 
$v \sin{i}=3.7\pm0.6$\,km\,s$^{-1}$ for \object{BD$-$21\,1074A}, and 
$T_{\rm eff}=3600\pm100$\,K, $\log{g}=4.5\pm0.1$\,dex, 
and $v \sin{i}=4.9\pm1.0$\,km\,s$^{-1}$ for \object{BD$-$21\,1074B} (see Table~\ref{tab:stellar_parameters}). 
The errors include the 1$\sigma$ standard deviation on the average 
and the typical errors of the PASTEL astrophysical parameters added 
in quadrature. We note that we did not detect any evidence of the component C, neither in the stellar spectra nor in the cross-correlation functions due to its relative faintness, compared to the component B.

\subsection{Chromospheric activity}
\label{sec:chromospheric_activity}

As described in Sect. 3.3, the HARPS spectra encompass a number of chromospheric lines among which most of the members of the 
Balmer series, the \ion{Ca}{ii} H \& K lines, the \ion{Na}{i} D$_{1,2}$ lines, and the \ion{He}{i} D$_{3}$ are present. Some of these lines appear in the 
spectra of \object{BD$-$21\,1074A} and \object{BD$-$21\,1074B} as pure emission features.    For a comparison with the literature, the average values of the EW$_{\rm H\alpha}$ measured on our HARPS spectra are 2.2 and 3.8\,\AA\,\, for the primary and secondary components, respectively.  However, for a proper evaluation of the chromospheric activity level,
we have considered the photospheric contribution in the aforementioned diagnostics.

 
Thus, we derived the chromospheric losses by means of the ``spectral subtraction'' technique, which is based on the comparison between the target spectrum, and a  non-active template, which can be a synthetic spectrum or that of slowly-rotating, low-activity star. The difference between the observed and the template 
removes the photospheric lines and leaves the net chromospheric line emission, from which we can get the total radiative losses in the line (see, e.g., Barden  \cite{Barden85}; Frasca \& Catalano  \cite{Frasca94}; Stelzer et al.  \cite{Stelzer13}). The residual equivalent width ($EW$) of the lines was measured by integrating the emission profile in the 
difference spectrum, while the error ($\sigma_{\rm EW}$) was estimated by multiplying the integration range by the photometric error on each point. 
The latter was evaluated by the standard deviation of the observed fluxes on the difference spectrum in two spectral regions close to the line. 
We made the measurements of the residual emission with both synthetic BT-Settl spectra (Allard et al.  \cite{Allard11}) and low-activity stellar spectra, obtaining nearly the same results. 
As a reference star, we used  the HARPS spectrum of \object{HD\,36395} for both targets (see Sect.\,4.2). Table~\ref{tab:equivalent_widths} reports the net equivalent widths of the H$\alpha$, H$\beta$, H$\gamma$, H$\delta$, H$\epsilon$, \ion{He}{i} $\lambda$5876, and \ion{Ca}{ii} H\&K lines. 

For the \ion{Ca}{ii} H \& K lines, the measurement of the residual emission is a more complex task due to the very broad photospheric absorption 
wings of these lines and to the huge crowding of absorption lines, which prevents a safe identification of a continuum (or a pseudo-continuum) 
in this spectral region. We thus used the method proposed by Frasca et al. (\cite{Frasca00}), which basically consists of a flux calibration of the segment of spectrum 
around the \ion{Ca}{ii} lines. The net equivalent width of the \ion{Ca}{ii} H \& K lines was thus referred to the average flux evaluated in two 10-\AA\  windows 
centered at  3910 and 4010\,\AA, respectively.  As non-active templates for subtracting the minimum photospheric flux, we used the BT-Settl spectra
with solar metallicity, $\log g=4.5$, and $T_{\rm eff}=3700$\,K (3600\,K), for the primary (secondary) component.

\begin{table*}
\caption{Radial velocities and equivalent widths of H$\alpha$, H$\beta$, H$\gamma$, H$\delta$, H$\epsilon$, 
\ion{He}{i} $\lambda$5876 \AA, \ion{Ca}{ii} H$\&$K, and Li $\lambda$6708 \AA~lines measured for each spectrum.}
\label{tab:equivalent_widths}
\begin{center}  
\begin{tabular}{ccccccccccc}  
\hline
\hline
\noalign{\smallskip}
$JD$ &  $V_{\rm rad}$ & $EW_{\rm H\alpha}$ & $EW_{\rm H\beta}$ & $EW_{\rm H\gamma}$ & $EW_{\rm H\delta}$ & $EW_{\rm H\epsilon}$ & $EW_{\rm He\,5876}$ & $EW_{\rm \ion{Ca}{ii}\,H}$ & $EW_{\rm \ion{Ca}{ii}\,K}$ & $EW_{\rm Li}^a$  \\ 
(+2\,450\,000) &   (kms$^{-1}$)	  &   (\AA)    &   (\AA)    &	(\AA)	 &   (\AA)    &   (\AA)     &	(\AA)	 &   (\AA)    &   (\AA)    &	(m\AA)    \\ 
\noalign{\smallskip}
\hline
~\\
\multicolumn{11}{c}{BD$-$21\,1074A}\\
6302.148 & 22.152$\pm$0.002& 2.7$\pm$0.2& 2.2$\pm$0.1& 2.0$\pm$0.2& 1.8$\pm$0.1& 2.1$\pm$0.2& 0.15$\pm$0.03& 6.0$\pm$0.3& 8.3$\pm$1.0&31$\pm$5\\ 
6305.141 & 22.022$\pm$0.002& 2.9$\pm$0.2& 2.3$\pm$0.1& 2.1$\pm$0.1& 2.0$\pm$0.1& 2.2$\pm$0.3& 0.13$\pm$0.03& 6.2$\pm$0.3& 8.5$\pm$0.6&32$\pm$5\\ 
\noalign{\smallskip}
\hline
~\\
\multicolumn{11}{c}{BD$-$21\,1074B}\\
6302.137 & 23.195$\pm$0.003& 5.4$\pm$0.3& 4.7$\pm$0.2& 4.7$\pm$0.2& 3.8$\pm$0.3& 3.4$\pm$0.6& 0.32$\pm$0.04& 7.9$\pm$0.6& 10.3$\pm$0.6&19$\pm$5\\ 
6303.129 & 23.194$\pm$0.003& 5.4$\pm$0.2& 4.8$\pm$0.2& 4.8$\pm$0.2& 3.6$\pm$0.2& 3.7$\pm$0.4& 0.28$\pm$0.06& 8.2$\pm$0.5& 11.5$\pm$0.8&17$\pm$5\\ 
6305.129 & 22.769$\pm$0.004& 6.0$\pm$0.1& 5.2$\pm$0.2& 5.1$\pm$0.2& 3.6$\pm$0.2& 3.7$\pm$0.5& 0.38$\pm$0.04& 8.1$\pm$0.5& 11.5$\pm$0.7&16$\pm$5\\ 
\noalign{\smallskip}
\hline
\end{tabular}
\tablefoot{ The lithium EWs were measured through direct integration using the IRAF {\sc splot} task. Errors in lithium EWs   represent the standard deviations of three EW measurements. \\
$^a$The contribution of the \ion{Fe}{i} line at $\lambda$6707.44\,\AA~
to the Li line was subtracted using the empirical correction of Soderblom et al. (\cite{Soderblom93}).}
\end{center}  
\end{table*}

Starting from the equivalent width of each line, $EW$,  we calculated the chromospheric line flux at the star surface as
\begin{eqnarray}
F_{\rm line} & = & F_{\rm cont}EW, 
\end{eqnarray}

and the activity R$^{\prime}$ index as
\begin{eqnarray}
R^{\prime}_{\rm line} = L_{\rm line} / & L_{\rm bol},
\end{eqnarray}

where {\noindent $F_{\rm cont}$ is the continuum surface flux in erg\,cm$^{-2}$s$^{-1}\AA^{-1}$ at the line  wavelength.} 
The latter was measured in the two sides of each line for the BT-settl  spectra (Allard et al.  \cite{Allard11}), and the value 
corresponding to our targets was taken by interpolating at the effective temperature and surface gravity of the targets reported in Table 4.
The fluxes and the R$^{\prime}$ index values were calculated for the H$\alpha$, H$\beta$, H$\epsilon$, and \ion{Ca}{ii} H\&K lines, and their values are listed in Table 5.

\begin{table*}
\caption{Chromospheric line fluxes and R$^{'}$ index.}
\centering
 \begin{tabular}{lcccccccccc}
  \hline\hline
  \noalign{\smallskip}
$JD$  &  $F_{H\alpha}$ &  logR$^{'}_{H\alpha}$         &   $F_{H\beta}$ &  logR$^{'}_{H\beta}$   & $F_{H\epsilon}$   &   logR$^{'}_{H\epsilon}$ &  $F_{\ion{Ca}{ii} H}$  &  logR$^{'}_{\ion{Ca}{ii} H}$    &  $F_{\ion{Ca}{ii} K}$   &  logR$^{'}_{\ion{Ca}{ii} K}$ \\  			     
\scriptsize{(+2\,450\,000)}  &  \multicolumn{10}{c}{} \\
  \noalign{\smallskip}
  \hline
  \noalign{\medskip}
  \multicolumn{10}{c}{BD$-$21 1074A} \\
  \noalign{\smallskip}
 6302.148  &  1.89$\pm$0.35   & $-3.76$   &  1.10$\pm$0.33   & $-3.99$ & 0.24$\pm$0.06  & $-4.66$ & 0.68$\pm$0.14  & $-4.20$ & 0.77$\pm$0.18  & $-4.15$ \\
 6305.142  &  2.03$\pm$0.38   & $-3.73$   &  1.15$\pm$0.35   & $-3.98$ & 0.25$\pm$0.06  & $-4.63$ & 0.70$\pm$0.15  & $-4.19$ & 0.78$\pm$0.17  & $-4.14$ \\
  \noalign{\smallskip}
   \hline
 \noalign{\medskip}			    
  \multicolumn{10}{c}{BD$-$21 1074B} \\
  \noalign{\smallskip}
 6302.135    &  3.13$\pm$0.67    &  $-3.48$ & 1.64$\pm$0.71  &  $-3.76$ & 0.29$\pm$0.09  & $-4.52$ & 0.69$\pm$0.19  & $-4.14$ &  0.73$\pm$0.20 & $-4.12$ \\
 6303.130    &  3.13$\pm$0.66    &  $-3.48$ & 1.68$\pm$0.72  &  $-3.75$ & 0.32$\pm$0.09  & $-4.47$ & 0.71$\pm$0.19  & $-4.13$ &  0.81$\pm$0.23 & $-4.07$ \\
 6305.129    &  3.48$\pm$0.72    &  $-3.44$ & 1.82$\pm$0.78  &  $-3.72$ & 0.32$\pm$0.10  & $-4.47$ & 0.70$\pm$0.19  & $-4.13$ &  0.81$\pm$0.22 & $-4.07$ \\
   \noalign{\smallskip}
  \hline
\end{tabular}
\label{tab:fluxes}
\begin{list}{}{}		         	                	                 
\item[] Notes. The errors include both the $EW$ error as quoted in Table 3 and the continuum-flux error estimated from the $T_{\rm eff}$ and 
$\log g$ errors evaluated with ROTFIT and reported in Table 4. Fluxes are expressed in $10^6$erg\,cm$^{-2}$\,s$^{-1}$. Errors on R$^{'}$
index are on the order of $^{+0.09}_{-0.11}$.
\end{list}
\end{table*}

 Both stars display high levels of chromospheric activity.
More specifically, component B appears to be slightly more active than A in all lines.
Both components  displays also an H$\alpha$ flux in excess with respect to \ion{Ca}{ii}, such as the stars in the upper branch of the
flux-flux diagram of Fig. 4 (Martinez-Arnaiz et al.  \cite{Martinez-Arnaiz11}).   This behavior is more evident in the B component.  
The fluxes we measured are also close to the maximum values found by Stelzer et al. ( \cite{Stelzer13}) in a sample of Class\,III stars belonging
to three young associations. 

The flux ratios also provide some insight into the chromospheric properties and the physical conditions of the emitting material.
The Balmer decrement,  $F_{H\alpha}/F_{H\beta}$, is lower than two for both stars, like the minimum values observed 
by Stelzer et al. ( \cite{Stelzer13}) for stars of comparable $T_{\rm eff}$, and it is close to the solar plage values (see their Fig. 13). 
More interesting is the behavior of the flux ratio $F_{\ion{Ca}{ii} K}/F_{H\alpha}$, which is about 0.25--0.40 for both components and nearly
equal to the value measured by  Stelzer et al. (\cite{Stelzer13})  for three members of TW~Hya association (8 Myr; Torres et al. 2008) of similar temperature and resembling the ``saturated'' stars.


\subsection{Stellar axis inclination}
 
\begin{figure}
\begin{center}
\begin{minipage}{9cm}
\centerline{
\includegraphics[width=6.cm,angle=90]{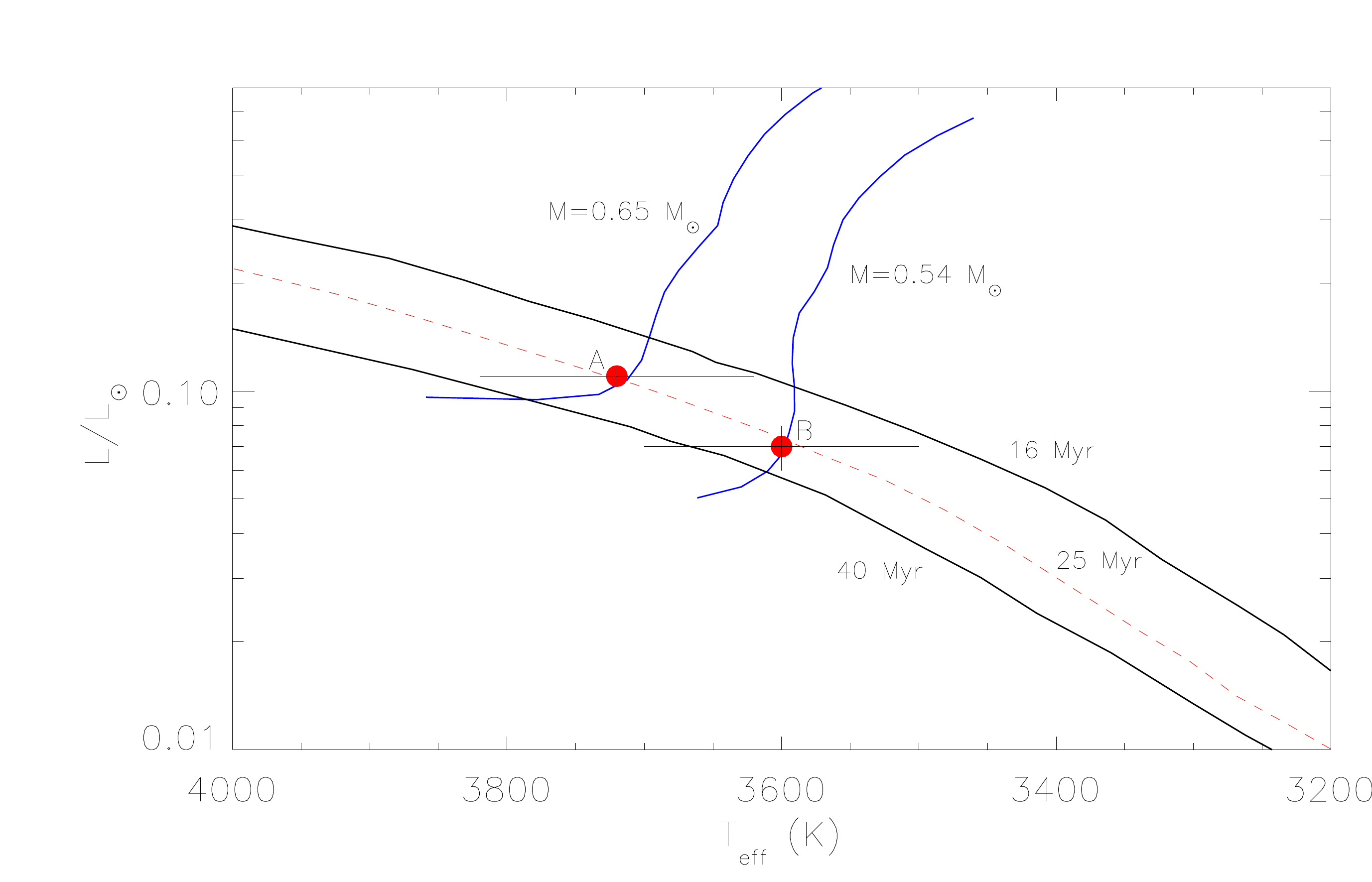}
}
\end{minipage}
 \caption{HR diagram of the components A and B with overplotted isochrones and mass tracks from Baraffe et al. (\cite{Baraffe98}).}
 \label{HR}
\end{center}
\end{figure}

We can use the effective temperature and luminosity to derive the stellar radii of both components and the inclination of the rotation axes. Assuming a distance d = 19.25 pc (Riedel et al.  \cite{Riedel14}), the brightest observed magnitudes V$_{\rm A}$ = 10.28 mag (Reid et al.  \cite{Reid04}) and V$_{\rm B}$ = 11.01 (Jao et al.  \cite{Jao03})\footnote{Since the flux from the component C is about 2\% of that from component A, the magnitude variation observed by Jao et al. (\cite{Jao03})
must be attributed to either A or B. Since we know the presumed brightest/unspotted magnitude
of component A, we can infer a limit for the presumed brightest/unspotted magnitude of component B (V=11.01\,mag).}, and bolometric corrections BC$_{\rm A}$ = $-$1.70\,mag and BC$_{\rm B}$ = $-$1.92 (Pecaut \& Mamajek  \cite{Pecaut13}), we infer the bolometric magnitudes Mb$_{\rm A}$ = 7.16$\pm$0.05\,mag and Mb$_{\rm B}$ = 7.67$\pm$0.05\,mag,  the luminosities L$_{\rm A}$ = 0.11$\pm$0.01\,L$_\odot$ and L$_{\rm B}$ = 0.07$\pm$0.01\,L$_\odot$, and the stellar radii R$_{\rm A}$ = 0.80$\pm$0.05\,R$_\odot$ and R$_{\rm B}$ = 0.68$\pm$0.05\,R$_\odot$. A lower limit to the stellar radius can also be inferred from the projected rotational velocity and the rotation period. We find R$_{\rm A} > 0.84$\,R$_\odot$ and  R$_{\rm B} > 0.69$\,R$_\odot$. These values are about 5\% larger than those derived from T$_{\rm eff}$ and luminosities but are still within the uncertainties deriving from $v \sin{i}$. From these values, we infer the inclinations of the stellar rotation axis  $i_A$ $\simeq$ 60$^\circ$ and  $i_B \simeq$ 50$^\circ$.\\
 The rotation period P = 13.6d of component A, inferred from the ASAS, SWASP data and the LS analysis, could still give a value of $\sin{i} < 1$, when the uncertainties on $v \sin{i}$, P, and R are considered. However, such a large rotation period is not consistent with the age and the start-disc locking typical duration. As it will be discussed in Sect.\,5, this large period would imply that component A  is still disc-locked after 21 Myr of life or has just started its spinning up. 
In Fig.\,\ref{HR}, we compare T$_{\rm eff}$ and luminosities of the components A and B with a set of mass tracks and isochronones taken from Baraffe et al. (\cite{Baraffe98}) for solar metallicity. We see that both components  are best fitted by the same isochrone of  about 25 Myr (the uncertainty on T$_{\rm eff}$ put the age in the range 16--40 Myr) and by mass tracks of  M$_A$ = 0.65$\pm$0.05M$_{\odot}$ and M$_B$ = 0.54$\pm$0.05M$_{\odot}$. The mass of the component C is expected to be  $\le$0.2M\,$_\odot$ and corresponds  to a spectral type M5 or later (see Sect. 4.3).\\
The object \object{BD$-$21\,1074} is one of the targets of the SACY survey (Search for Associations containing Young Stars; Torres et al.  \cite{Torres06}) that is proposed to be a member of the young \object{$\beta$ Pictoris} moving group by Torres et al. (\cite{Torres08}) to which they assign an age of 10 Myr. The membership of \object{BD$-$21\,1074} has been recently confirmed by Malo et al. (\cite{Malo13}) and Riedel et al. (\cite{Riedel14}). Very recently, Binks \& Jeffries (\cite{Binks14}) found the lithium depletion boundary age for this association to be significantly older, which is  21$\pm$4 Myr. 
We also find the age of this member of \object{$\beta$ Pic} to be more consistent with the age of 21 Myr quoted by Binks \& Jeffries (\cite{Binks14}) than that from Torres et al. (\cite{Torres08}).

\section{Discussion}
Our photometric and spectroscopic analyses  allowed us to characterize the rotational, physical, and magnetic activity properties of the two brightest components of the system. A larger aperture telescope with sub arcsec spatial resolution is needed to characterize the faintest component C. The brightest components are two M1.5 + M2.5 Pre Main Sequence dwarfs, whose radii are still larger and surface gravity still smaller than their MS counterparts. Both stars are active; the secondary being slightly more active than the primary component (see Table 3), which agrees with its deeper convection zone and its faster rotation rate. Moreover, both components show clear evidence of variability at the photospheric level (from the light rotational modulation) and strong chromospheric activity  (from the Balmer and $\ion{Ca}{ii}$ H\&K emission lines). This level of magnetic activity is also manifested by the observation of a few flare events and by the X-ray coronal emission.\\
\indent
What makes this system particularly interesting is that it consists of three low-mass stars physically bound with the same age and initial chemical composition but different rotation periods. \\
This means that either their initial rotation periods or their rotational evolution have been significantly different. The brightest component A has a rotation period of P = 9.3\,d and is at a distance of about 160 AU from the components BC. The component B has P = 5.4\,d as rotation period and is at a distance of about 15 AU from the component C.
  During the first Myrs of life we know that low-mass stars can be surrounded by an accretion disc. The primordial disc frequency decreases exponentially with a time scale of about 5 Myr, and discs are no longer detected for ages older than 10--20 Myr (Ribas et al.  \cite{Ribas14}). However, it is unlikely that the disc-locking be effective for star older than 10 Myr when accretion signatures are no longer detected (Hern\'andez et al.  \cite{Hernandez07},  \cite{Hernandez08}; Frasca et al.  \cite{Frasca14}).   Such discs, owing to a magnetic interaction with its host star, give rise to a magnetic locking that prevents the star outer envelope to spin-up notwithstanding the ongoing radius gravitational contraction (see, e.g., M\'enard \& Bertout  \cite{Menard99}). Only after the disc dispersal, the star is free to increase its rotational velocity. The difference of rotation period we measured in this framework cannot be ascribed to the difference in mass (0.65 M$_\odot$ versus 0.55 M$_\odot$). The dissipation timescales are about the same for substellar objects and solar-mass stars  (see Williams \& Cieca  \cite{Williams11}, and references therein). They are significantly shorter only for higher masses.   This period difference may indicate that either the initial rotation periods were significantly different or the star-disc locking of the component A has been lasting longer than that for the B component. It is reasonable to suppose that the vicinity between the components B and C has facilitated the circumstellar/circumbinary disc disruption and its dispersal by allowing the B and C components to start their spin up at earlier time. This implies that components A and BC will reach the ZAMS with different rotation periods. Assuming that the mass difference does not play a significant role, then the different environment of the component A with respect to the component B, which has a close-by 'perturber' (i.e., the component C), has determined the rotation period difference. This can be considered as one effective cause of the rotation period dispersion observed in an open cluster among stars sharing similar mass, age, and chemical composition, and is different for the presence of a disc. Different rotation evolution models (see, e.g., Spada et al.  \cite{Spada11}; Gallet \& Bouvier  \cite{Gallet13}, just to mention the most recent) all describe very similarly the evolution of normalized (to the solar value) angular velocity in the age range from about 5 to 30 Myr. The angular  velocity  linearly increases in  log($\Omega/\Omega_\odot$)-log(age) scale. This allows us to estimate the age difference corresponding to the rotation period difference between 
components A and B. 
To do that, we need to make some assumption on the initial values of the rotation periods P$_{in}$ of both components A and B during the disc-locking phase. Thereafter, we   consider the prescriptions given in the Gallet \& Bouvier (\cite{Gallet13}) 
model as our guideline. 
The rotation period distribution of the \object{Orion Nebula Cluster} members at about 1 Myr can be considered as representative of the period distribution during the disc-locking phase. We thus retrieved the rotation periods in the same 
mass range as our components from Herbst et al. (\cite{Herbst02}), which are 0.5-0.7 M$_\odot$.
The period distribution is found to be double peaked with the longer periods ranging from 8 to 14 days. We then assume that our components A and B have their  P$_{in}$ within this range  during the disc-locking phase.  \\

\begin{figure}
\begin{center}
\begin{minipage}{9cm}
\centerline{
\includegraphics[width=6.cm,angle=90]{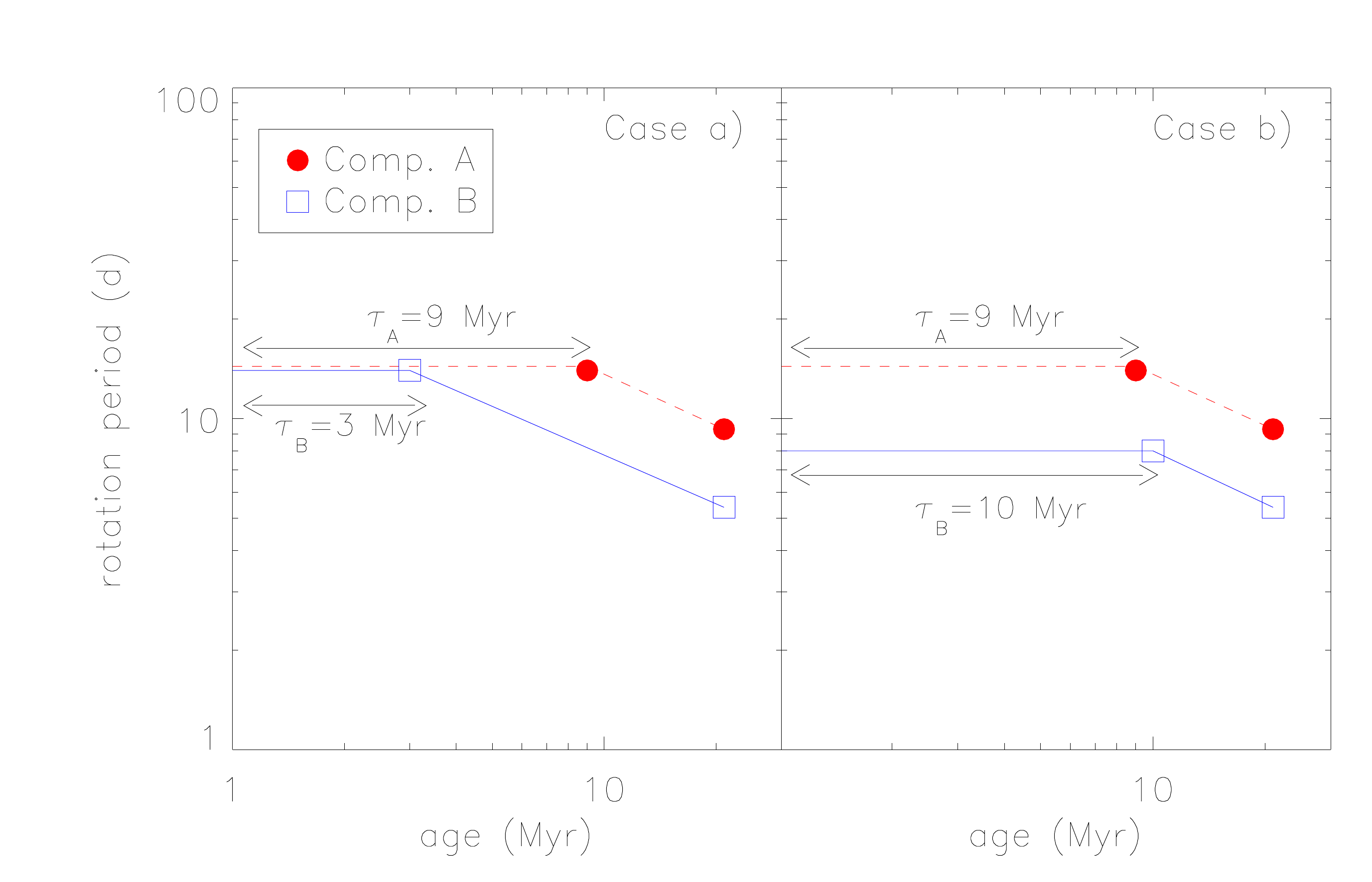}
}
\end{minipage}
 \caption{Qualitative description, according to the model of Gallet \& Bouvier (\cite{Gallet13}), of the inferred disc lifetime in two extreme cases: (a) same initial rotation periods of component A and B and (b) maximum differences of their initial rotation periods.}
 \label{model}
\end{center}
\end{figure}
Now, we can consider the two extreme cases: 
 a) no difference in initial 
periods (i.e., both components with P$_{in}$ =14d); b)  large difference 
in initial periods (i.e., P$_{in_A}$=14\,d and P$_{in_B}$=8\,d).\\  
In case (a), components A and B have the same initial rotation period P$_{in}$=14\,d (normalized angular velocity $\Omega/\Omega_\odot$=1.80), whereas the present values are P$_{\rm  A}$=9.3\,d and P$_{\rm  B}$=5.4\,d ($\Omega_{\rm A}/\Omega_\odot$=2.72 and $\Omega_{\rm B}/\Omega_\odot$=4.69, where $\Omega_\odot=2.87 \times 10^{-6}$  s$^{-1}$). According to the  Gallet \& Bouvier (2013) model  (see also the left panel of Fig.\,\ref{model}), it takes about 11 Myr for component A to spin from P$_{in}$ =14\,d to P=9.3\,d (from $\Omega_{\rm A}/\Omega_\odot$=1.8 to $\Omega_{\rm A}/\Omega_\odot$=2.72), whereas it takes about 17 Myr for component B to spin from P$_{in}$ =14\,d to P=5.4\,d (from $\Omega_{\rm B}/\Omega_\odot$=1.8 to $\Omega_{\rm B}/\Omega_\odot$=4.69). Therefore, we infer that component A has experienced about 9 Myr of disc-locking, whereas component B has experienced only 3 Myr of disc-locking.  Assuming 10 Myr as maximum duration of disc-locking, we note that the initial rotation period of component A could not be shorter than 13\,d. \\
In case (b), component A has an initial rotation period P$_{in}$=14\,d ($\Omega_{\rm A}/\Omega_\odot$=1.80), and component B has an initial rotation period P$_{in}$=8\,d ($\Omega_{\rm B}/\Omega_\odot$=3.50).
According to the model (see also the right panel of Fig.\,\ref{model}), it takes about 11 Myr for component A to spin from P$_{in}$=14\,d to P=9.3\,d again. Whereas it takes about 10 Myr for component B to spin from P$_{in}$=8\,d to P=5.4\,d (from $\Omega_{\rm B}/\Omega_\odot$=3.50 to $\Omega_{\rm B}/\Omega_\odot$=4.69), it is longer if B had P$_{in}$ $>$ 8\,d.\\
To summarize, the disc-locking phase of component B, which can range from 3 to 10 Myr, is generally shorter than that of component A,  which can range from 9 to 10 Myr, either if we assume same initial rotation period P$_{in}$ =14\,d or P$_{in{\rm B}}$ in the range 14--8d. The only case where disc-locking phases are comparable is case (b), when component A and B have  a large difference  in their initial periods.\\
On the basis of the considerations given above, we infer that the presently observed difference of rotation periods between component A and B can be attributed entirely to different initial rotation periods in the only case if they had  a large difference of initial rotation periods (P$_{in{\rm A}}$=14\,d and P$_{in{\rm B}}$=8\,d). In all the other cases, differences in the disc lifetime must be invoked with the disc lifetime of component B shorter than that of component A. \\
For instance, we note that our conclusions do not change significantly if we assume an age of 10 Myr for the $\beta$ Pic association.
In this hypothesis, the disc-locking phase of component A must be longer than 5 Myr with P$_{in_A}$ ranging from 14\,d to 11\,d, and the disc-locking phase of component B must be longer than about 1 Myr. This implies that there is a slightly larger range of initial values for component A for attributing the observed rotation period difference to only different initial values.\\
 
As anticipated, we are carrying out similar studies on other multiple systems with different ages and rotation periods and with perturbing components that are at different distances from the target stars, in which we expect evidence from rotation of shortened disc lifetimes. The results will be presented in subsequent papers.   To enlarge the sample of well-studied cases, it is important also to address somewhat conflicting results from earlier studies, such as in the case of Bouvier et al. (\cite{Bouvier97}), who found no difference in the rotational velocities distribution of single stars with respect to visual binaries in a sample of Pleiades members.   \\
Another relevant aspect of our analysis is that  no planets are expected to have been formed in case the disc dispersal is enhanced, unless they formed very rapidly by gravitational instability mechanism. This circumstance makes these stars particular interesting and worth investigating for planets searches. For instance, we note that Riedel et al. (\cite{Riedel14}) investigated the B and C components by collecting observations with the Hubble Space Telescope Fine Guidance Sensor Interferometer. Although observations clearly resolve the B and C component; however, they could neither wrap the orbit nor find convergence of the orbit fit. This may suggest the presence of an additional body, making this system certainly worth of additional investigation.

\begin{table}
\caption{Stellar parameters derived with ROTFIT.}
\label{tab:stellar_parameters}
\centering
\begin{tabular}{ccccc}  
\hline
\hline
\noalign{\smallskip}
Object & Sp. Type & $T_{\rm eff}$ & $\log{g}$  &  $v\sin{i}$    \\ 
       &          &   (K) 	  &  (dex)     &  (km s$^{-1}$)  \\ 
\noalign{\smallskip}
\hline
\noalign{\smallskip}
\object{BD$-$21\,1074A} & M1.5$\pm$0.5 & 3720$\pm$100 & 4.7$\pm$0.1 & 3.7$\pm$0.6\\
\object{BD$-$21\,1074B} & M2.5$\pm$0.5 & 3600$\pm$100 & 4.5$\pm$0.1 & 4.9$\pm$1.0\\
\noalign{\smallskip}
\hline
\end{tabular}
\end{table}

\begin{figure}
\begin{minipage}{10cm}
\includegraphics[width=70mm,height=90mm, angle=90]{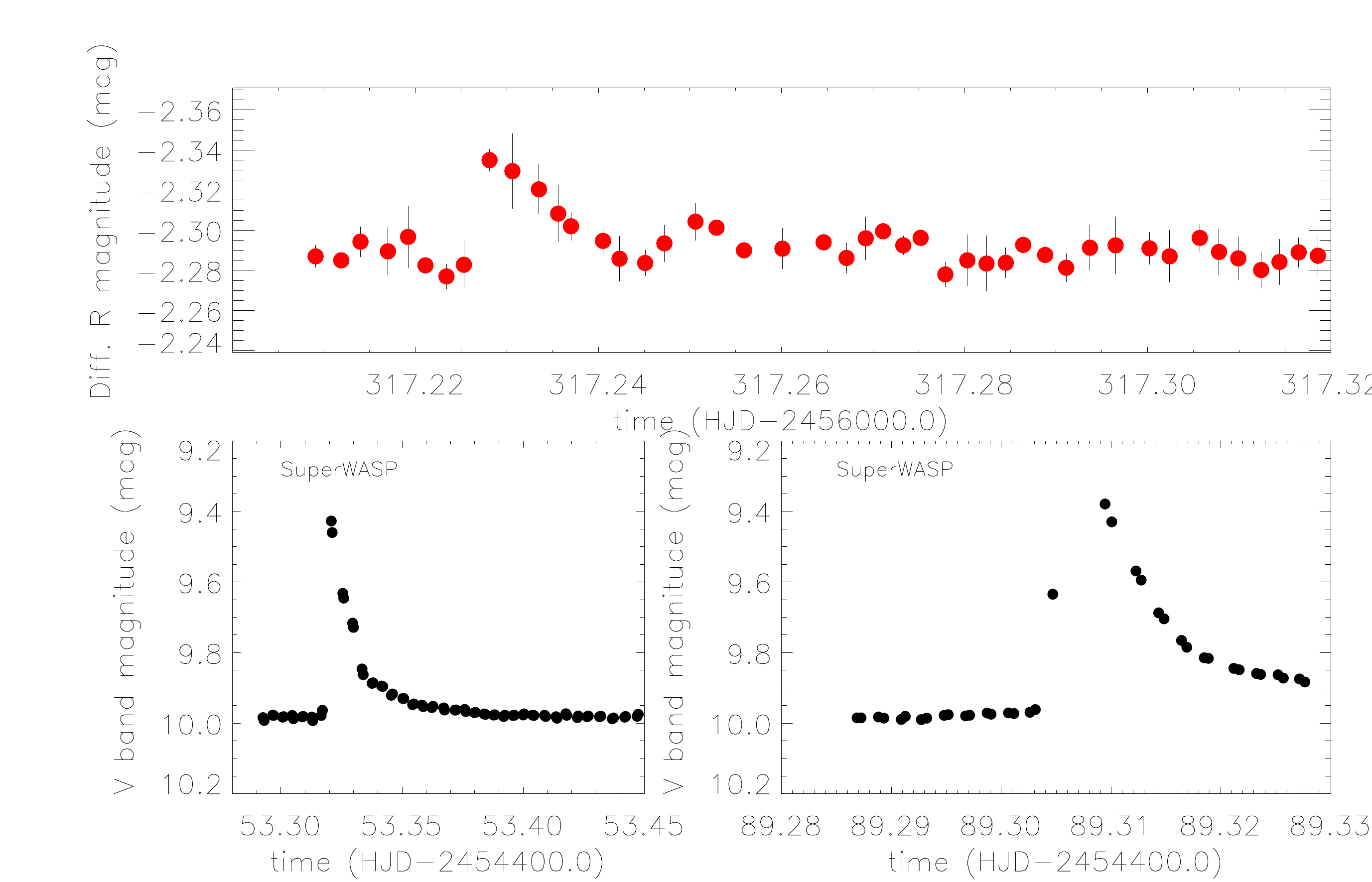}
\end{minipage}
\caption{\label{flare}Examples of flares detected on the component BD$-$211074A in the  R-band during the observation run carried out by us (top panel) and in the V band from the SuperWASP time series for the unresolved system (bottom panels).}
\vspace{0cm}
\end{figure}

\section{Conclusions}
Our photometric and spectroscopic observations of the visual triple system \object{BD$-$21\,1074} have allowed us to derive the stellar rotation periods of the A and B components, P = 9.3\,d and P = 5.4\,d, respectively, and  to confirm their high level of magnetic activity, where the M2.5 component is slightly more active than the M1.5 component. Our measured projected rotational velocities have allowed us to infer the inclinations of the stellar rotation axis, which   are similar  $i$ $\simeq$ 60$^\circ$ and  $i$ $\simeq$ 50$^\circ$ for components A and B, respectively.   Components A and B exhibit significant difference between their rotation periods. Since they have a similar mass, age, and initial chemical composition,   and only the largest difference of initial rotation periods can alone explain the observed period difference,  we think that differences in the lifetime of their primordial discs must also be invoked and, consequently, the different duration of their disc-locking phase to explain the presently observed rotation period difference.   The presence of the component C, which is closer to B (15 AU) than to A (160 AU),  has likely enhanced the dispersal of the primordial disc of component B, making it free to spin up its rotation earlier than A. Actually, comparing the period difference with models of angular momentum evolution, the duration of the  star-disc locking phase of component B ranges from 3 to 10 Myr, whereas that of component A ranges from 9 to 10 Myr.  
Our hypothesis of the enhanced disc dispersal plays against the eventual formation of planets by core accretion, allowing only gravitational instability as possible mechanism. This circumstance in the case of future planet detections makes this system particularly interesting in light of understanding the planet formation mechanisms. \\
  In the specific case of  \object{BD$-$21\,1074}, the scenario of an enhanced disc dispersal seems to be favored with respect to the scenario of a large difference of initial rotation periods, which is an explanation of the currently observed rotation period difference between components A and B. However, this is just one system whose behavior may be peculiar. A statistically meaningful sample of such systems must be investigated before we can establish which scenario is actually more likely.
 Such systems will be investigated in forthcoming papers.

\section*{Acknowledgements}
The extensive use  of the SIMBAD  and ADS  databases  operated by  the  CDS center,  Strasbourg,
France,  is gratefully  acknowledged. We thank the Super-WASP consortium for the use of their public archive in this research.
The Authors  thank the Referee for very helpful comments.

\end{document}